\documentclass[a4paper,12pt]{article}

\pdfoutput=1
\usepackage{amsfonts,amsthm,amsmath,amssymb,graphicx}
\usepackage{braket}
\usepackage{amsmath}
\usepackage{url}
\usepackage{graphicx,color}

 \def\ep{{\epsilon}}

 \def\frac#1#2{{#1\over #2}}

 \def\s{\sqrt}

\def\be{\begin{equation}}
\def\ee{\end{equation}}
\def\ba{\begin{eqnarray}}
\def\ea{\end{eqnarray}}
\numberwithin{equation}{section}

 \def\de{\partial}

 \def\lr{\leftrightarrow}
 \def\f {\frac}
 \def\ti{\tilde}

 \def\ddd{\cdot\cdot\cdot}
 \def\no{\nonumber \\}

 \def\la{\langle}
 \def\lb{\rangle}
 \def\ep{\epsilon}

 \def\vp{\varphi}
\usepackage{hyperref}
\begin{document}

\begin{titlepage}
\thispagestyle{empty}

\begin{flushright}
YITP-18-28
\\
IPMU18-0061
\\
UK-18-03
\\
\end{flushright}

\bigskip

\begin{center}
\noindent{{ \large \textbf{Path-Integral Complexity for Perturbed CFTs}}}\\
\vspace{2cm}

Arpan Bhattacharyya$^{a}$, Pawel Caputa$^{a}$, Sumit R. Das$^{b}$, \\
Nilay Kundu$^{a}$, Masamichi Miyaji$^{a}$ and Tadashi Takayanagi$^{a,c}$
\vspace{1cm}

{\it
$^{a}$Center for Gravitational Physics, \\
Yukawa Institute for Theoretical Physics,
Kyoto University, \\
Kyoto 606-8502, Japan\\
$^{b}$ Department of Physics and Astronomy, \\
University of Kentucky, Lexington, KY 40506, U.S.A. \\
$^{c}$Kavli Institute for the Physics and Mathematics of the Universe,\\
University of Tokyo, Kashiwa, Chiba 277-8582, Japan\\
}

\vskip 2em
\end{center}

\begin{abstract}
In this work, we formulate a path-integral optimization for two dimensional conformal field theories perturbed by relevant operators. We present several evidences how this optimization mechanism works, based on calculations in free field theories as well as general arguments of RG flows in field theories. Our optimization is performed by minimizing the path-integral complexity functional that depends on the metric and also on the relevant couplings. Then, we compute the optimal metric perturbatively and find that it agrees with the time slice of the hyperbolic metric perturbed by a scalar field in the AdS/CFT correspondence. Last but not the least, we estimate contributions to complexity from relevant perturbations.
\end{abstract}

\end{titlepage}

\newpage

\section{Introduction}

The AdS/CFT correspondence provides a strong evidence that gravitational spacetimes can emerges from microscopic quantum theories \cite{Ma,GKP,Wi}. Usually, the extra radial coordinate in an AdS is identified with a length scale in the sense of a renormalization group flow \cite{dVV,Sk,DJ,HP,FLR,Lee}. One basic way to study the origin of the radial coordinate is to investigate quantum wave functionals in holographic CFTs at a fixed time. Recently, it was proposed that in the Euclidean path-integral description of wave functionals, the radial direction in the bulk AdS emerges by optimizing the path-integral for each quantum state \cite{Caputa:2017urj}.

The optimization procedure can be done as follows. First, imagine that we discretize the Euclidean path-integral into a lattice theory. Then we change the UV cut off scale (i.e. lattice spacing) locally \cite{MTW}. The position dependence of the cut off is systematically described by introducing a metric such that a unit area corresponds to a single lattice site \cite{Caputa:2017urj}. The basic requirement is that the wave functional in this non-trivial metric is proportional to the wave functional in a trivial metric with a normalization factor which {\em only} depends on the metric and couplings of the theory. As we will see, this requires position dependent couplings.
The optimization, which makes  computation of the path-integral most efficient, is realized by minimizing the overall normalization factor of the wave functional for the quantum state that we want (such as a CFT vacuum state) when we change the metric. The main new idea is that after such optimization procedure, the metric coincides with that of a time slice of holographic dual spacetime, e.g. a hyperbolic space for a CFT vacuum.

The idea of the path-integral optimization is closely related to the conjecture first proposed in \cite{Swingle} that MERA tensor networks \cite{MERA,TNR} may describe gravity duals and therefore explain the mechanism behind the AdS/CFT. Since then, there have been many improvements in constructions of tensor networks related to AdS/CFT, such as cMERA \cite{cMERA,NRT}, perfect tensor networks \cite{HAPPY} and random tensor networks \cite{HQ}.
Indeed, quantum states obtained from the path-integral optimization manifestly realize the surface/state correspondence \cite{MiTa}, which is the one of the most basic properties when we interpret holographic spaces as tensor networks. Moreover, the path-integral optimization approach has an advantage over tensor network approaches that we can take into account backreactions due to various excitations in a systematic way.

Recently, a very interesting and important problem in the AdS/CFT correspondence is to estimate complexity of quantum states. In two dimensional CFTs, the logarithm of the normalization factor of the wave functional is given by the Liouville action and can be naturally regarded as a path-integral definition of computational complexity for CFT states \cite{Caputa:2017urj}. Moreover, the minimization of the Liouville action in various setups leads to optimal metrics which coincide with time slices of dual AdS/CFT geometries linking the dynamics of gravity with complexity. A higher dimensional generalization of the path-integral complexity in CFTs has been given in \cite{Caputa:2017yrh}. Further support and explanations on the connection between the Liouville action and complexity were given in \cite{CzechC}.  On the gravity side, the holographic complexity in AdS/CFT have been proposed and formulated in \cite{SUR,Susskind,BrownSusskind1,Lehner:2016vdi}. Refer also to \cite{Chapman:2017rqy,JeMy,Molina-Vilaplana:2018sfn,KKS,HaMy} for evaluations of the circuit complexity in free quantum field theories.

The main aim of this paper is to extend this path-integral analysis of holographic emergent space to non-conformal examples. In particular we are interested in a relevant perturbation of a given CFT by a primary scalar operator and we generalize our optimization procedure to such examples. In order to keep the ground state wave functional for a perturbed CFT proportional to that in a flat metric, we allow the coupling constant of the perturbation to locally vary when we optimize the path-integral by tuning the metric. The position dependent coupling constant is naturally identified with the bulk scalar field. From this prescription we can determine the generalized path-integral complexity action for perturbed CFT that now depends on the metric as well as the couplings. Finally, we can estimate the metric after the optimization which corresponds to a perturbation of the hyperbolic space metric as expected from the AdS/CFT.

This paper is organized as follows. In section two, we explain a general formulation of our  path-integral optimization. In section three, we study the path-integral optimization of two dimensional CFTs and their relevant perturbations. In section four, we present examples of free scalar and fermion field theories. In section five, we give supports of the idea of the path-integral optimization by studying local RG flow. Moreover, we explicitly perform the path-integral optimization for two dimensional CFTs with relevant perturbations and calculate the path-integral complexity. In section six, we compare our results with those in AdS/CFT. In section seven, we summarize our conclusions and discuss future problems. In appendix A, we study the path-integral optimization of a massive scalar field theory to the leading order in the mass deformation.

\section{General Formulation of Path-Integral Optimization}

 Let us start by reviewing our conventions and a general procedure of the path-integral optimization for quantum field theories (QFTs) in $d$ dimensions \cite{Caputa:2017urj,MTW}. Consider a quantum field theory in the $d$ dimensional flat space $R^d$. We write the Euclidean time coordinate and $d-1$ dimensional space coordinates by $\tau$ and $x$, respectively. It is useful to define the coordinate $z=-\tau$, which is later, in comparison with holography, identified with that of the extra dimension in AdS. In the actual computations, we will set $d=2$ later. We represent all quantum fields in the QFT by $\vp(x,z)$ and represent all coupling constants in this QFT by $\lambda_0$.

The wave functional $\Psi$ of the QFT vacuum state can be computed by an Euclidean path-integral on a flat half space $-\infty<\tau<-\ep$ (or equally $\ep<z<\infty$) as
\be
\Psi_{g_0,\lambda_0}[\vp(x)]=\int \prod_{x}\prod_{\ep<z<\infty}[D\vp(x,z)]~e^{-S_{g_0,\lambda_0}[\vp]}~\prod_{x}\delta (\vp(x,\ep)-\vp(x)),\no  \label{phzo}
\ee
where $g_0$ denotes the flat space metric on which we perform the path-integration
\be
ds^2=\frac{1}{\mu \ep^2}(dx^2+dz^2).  \label{flatm}
\ee
Here $\sqrt{\mu}$ is an energy scale.
The infinitesimally small parameter $\ep$ is the UV cut off (lattice spacing) and we
introduce the rule that there is a single lattice site in a unit area of the background metric.
For the flat metric (\ref{flatm}), we can take a square lattice such that there are
$L^d/\ep^d$ lattice sites in the square $0<x,z<L$.  We also chose the end point of the path-integral to be $z=\ep$ instead of $z=0$ just for convenience. In the following we will make a choice of units, setting $\mu = 1$. The action with this metric and the coupling constant $\lambda_0$ is denoted by $S_{g_0,\lambda_0}[\vp]$.

\subsection{Optimization Procedure}
Now we are in a position to state our optimization procedure. We allow the metric $g_0$ and coupling constants $\lambda_0$ to non-trivially depend on the coordinates $(x,z)$. Namely, we write them as $g(x,z)$ and $\lambda(x,z)$ and impose the boundary conditions
\be
g(x,\ep)=g_0,\ \ \lambda(x,\ep)=\lambda_0. \label{pobc}
\ee
 At $z=\epsilon$ the metric remains same as the original  metric and thus the fields are not rescaled. So we need to impose this boundary condition which is same as what  is used for the original wavefunctional, so that after the rescaling of the metric (and the fields) the new wavefunctional will be same as the original one.

In general, the wave functional obtained from the Euclidean path-integral on this curved space with the position dependent coupling constants
\be
\Psi_{g(x,z),\lambda(x,z)}[\vp(x)]=\int \prod_{x}\prod_{\ep<z<\infty}[D\vp(x,z)]~e^{-S_{g(x,z),\lambda(x,z)}[\vp]}\prod_{x}\delta (\vp(x,\ep)-\vp(x)),\label{psigl}
\ee
differs from the original one $\Psi_{g_0,\lambda_0}[\vp(x)]$ (\ref{phzo}) in a nontrivial fashion. However, if we fine tune $g(x,z)$ and $\lambda(x,z)$, then we can find non-trivial functions $g(x,z)$ and $\lambda(x,z)$ which, in the path integral approach, give rise to  wave functionals proportional to the one computed with their boundary values $\Psi_{g(x,z),\lambda(x,z)}[\vp(x)]\propto \Psi_{g_0,\lambda_0}[\vp(x)]$. This means that they describe the same quantum state. More precisely, for these functions we can write
\ba
\Psi_{g(x,z),\lambda(x,z)}[\vp(x)]=e^{N[g,\lambda]-N[g_0,\lambda_0]}\cdot
\Psi_{g_0,\lambda_0}[\vp(x)].  \label{wsymw}
\ea
The optimization procedure can be completed by minimizing the normalization factor, or equivalently minimizing the functional $N[g,\lambda]$ with respect to $g(x,z)$ and $\lambda(x,z)$.
The position dependent metric and coupling constants which minimize $N[g,\lambda]$ are
written as $g_{min}$ and $\lambda_{min}$ and later we will suggest that they correspond to the metric and bulk fields on a time slice of AdS for holographic CFTs. In addition, we define the quantity $N[g_{min},\lambda_{min}]$ as the path-integral complexity (denoted by $C[\lambda_0]$) for the vacuum state in the QFT, given by the
wave functional $\Psi_{g_0,\lambda_0}[\vp(x)]$:
\ba
C[\lambda_0]\equiv\mbox{Min}_{g(x,z),\lambda(x,z)}\ N[g(x,z),\lambda(x,z)].
\label{pcdef}
\ea
It is also straightforward to extend the path-integral optimization to general excited states. This is because once  we have  a path-integral description of a quantum state, which we want to consider (e.g. inserting local operators in the middle of the path-integral) is given, in principle, the optimization by locally deforming the metric and coupling constants can be performed.

\subsection{Interpretation}

Before we go on to explicit examples, let us explain an intuitive idea behind our prescription. First, consider a numerical computation of path-integral to calculate $\Psi_{g_0,\lambda_0}[\vp(x)]$ (\ref{phzo}). We normally fine-grain both $z$ and $x$ coordinate for the metric (\ref{flatm}) such that the size of each cell is given by $\Delta z=\Delta x=\ep$. What we have in our mind here is that,  we perform the discretization of the path-integral in a way, such that each unit area square corresponds to a tensor $T_{a_1,a_2,..,a_n}$. For example, 
in Fig.~(\ref{fig:conformalTR}) we can interpret each square cell as a 
tensor with 4 indices ($n=4$). When two cells are attached along an edge,
we contract the corresponding indices. The path-integral is then 
approximated by all of such contraction of all these tensors.\footnote{Intuitively we can think that we have written the total path-integral as a product of transfer matrices and each of these tensors represents these transfer matrices.}
 However, if we think about the path-integral in an early time $\tau\to -\infty$, we do not need such fine grained information of the wave functional at that time as we path-integrate for a long time afterwards, as explained in \cite{MTW, Caputa:2017urj, Caputa:2017yrh}. This means that we can coarse-grain the cells in the past. In order to reproduce the correct wave functional  $\Psi_{g_0,\lambda_0}[\vp(x)]$, we need to reduce the amount of coarse-graining as the time evolves, ending up with the fined grained lattice at $z=\ep$. In terms of the tensors, we have to recombine the initial tensors 
and replace them by  fewer numbers  effective tensors. The righthand 
side of the Fig.~(\ref{fig:conformalTR})  shows this coarse grained 
geometry where again each of cells
is interpreted as a tensor with 5 indices ($n=5$). Again contractions of 
all indices of these tensors give the discretized path-integral.
This coarse-graining procedure can be systematically described by locally changing the metric and coupling constants as $g(x,z)$ and $\lambda(x,z)$ with the boundary condition (\ref{pobc}).

Next, we want to make the numerical computation of the path-integral the most efficient; the process which we call the optimization of the path-integral. Notice that here we do not want to change the dependence of the final wave functional  on the field configuration after the path-integration, which gives the constraint (\ref{wsymw}). We minimize the amount of algebraic computations in a lattice regularization to obtain the correct ground state wave functional. We argue this can be performed by minimizing the overall normalization of wave functional, given by $e^{N[g,\lambda]}$ in (\ref{wsymw}). This is because $N[g,\lambda]$ is an obvious measure which estimates the number of path-integral operations to obtain a given quantum state. This argument was also justified in \cite{CzechC} from the viewpoint of complexity of MERA tensor networks \cite{MERA}, which describe ground states of CFTs in two dimensions.

\section{Path-integral Optimization of 2D CFTs and Relevant Peturbations}

A class of QFTs where the optimization procedure is tractable is given by two dimensional conformal field theories \cite{Caputa:2017urj}. Here first we would like to briefly review an explicit optimization procedure for two dimensional CFTs, focusing on the vacuum state. Refer to \cite{Caputa:2017yrh} for more detailed computations as well as generalization to excited states. In addition, we will present an argument which provides an extra support of our procedure. Next we turn to the main aim of this work i.e. the path-integral optimizations of QFTs defined by relevant perturbations of two dimensional CFTs.

\subsection{Path-integral Optimization of 2D CFTs}

For conformal field theories, to optimize the path-integral, we only need to change the background metric locally. Therefore we can suppress the dependence on coupling constants $\lambda_0$.
In two dimensions, the metric can be chosen to be conformally flat,
\be
ds^2=e^{2\phi(x,z)}(dz^2+dx^2).  \label{metw}
\ee
As we already explained, we arrange a lattice regularization for any given $\phi(x,z)$
such that each lattice cell has the unit area in the metric $ds^2$ given by (\ref{metw}).
Thus, increasing $\phi$ means a fine-graining. In the original UV theory, we take the
flat metric (\ref{flatm}) i.e 
\be
e^{\phi(x,z)}=1/\ep \equiv e^{\phi_0}.  \label{UV}
\ee

The advantage of considering CFTs is that the Weyl rescaling (\ref{metw}) does not change the
quantum state owing to the local scale invariance. Therefore the vacuum wave functional
computed by (\ref{psigl}) coincides with the original one up to the normalization
factor given by an exponential of Liouville action \cite{Po,GM}
\ba
\Psi_{g=e^{2\phi}}[\vp(x)]=e^{S_L[\phi]-S_L[0]}\cdot
\Psi_{g=e^{2\phi_0}}[\vp(x)],  \label{prptdtt}
\ea
where $S_L$ is the Liouville action
\be
S_L[\phi]=\frac{c}{24\pi}\int dxdz \left[(\de_x\phi)^2+(\de_z\phi)^2+e^{2\phi} \right],
\ee
where $c$ stands for the central charge of the CFT. This way, the path-integral complexity (\ref{wsymw}) is simplify identified with the Liouville action $N[g=e^{2\phi}]=S_L[\phi].$  Note that if we restore units the potential term in the action will attain the standard form $\mu e^{2\phi}.$ 

The optimization can be achieved by minimizing the Liouville action $S_L[\phi]$ with the boundary condition (\ref{pobc}) at $z=\ep$ i.e.
\be
\phi(x,\ep)=\phi_0 =-\log \ep. \label{bclq}
\ee
Assuming the space coordinate $x$ is non-compact $-\infty<x<\infty$, this minimization leads to the following simple solution
\be
e^{\phi(x,z)}=\frac{1}{z}. \label{hypas}
\ee
Thus the optimized metric coincides with that of a two dimensional hyperbolic space H$_2$, which
agrees with the time slice of AdS$_3$ \cite{Caputa:2017urj}.

\begin{figure}
  \centering
  \includegraphics[width=9cm]{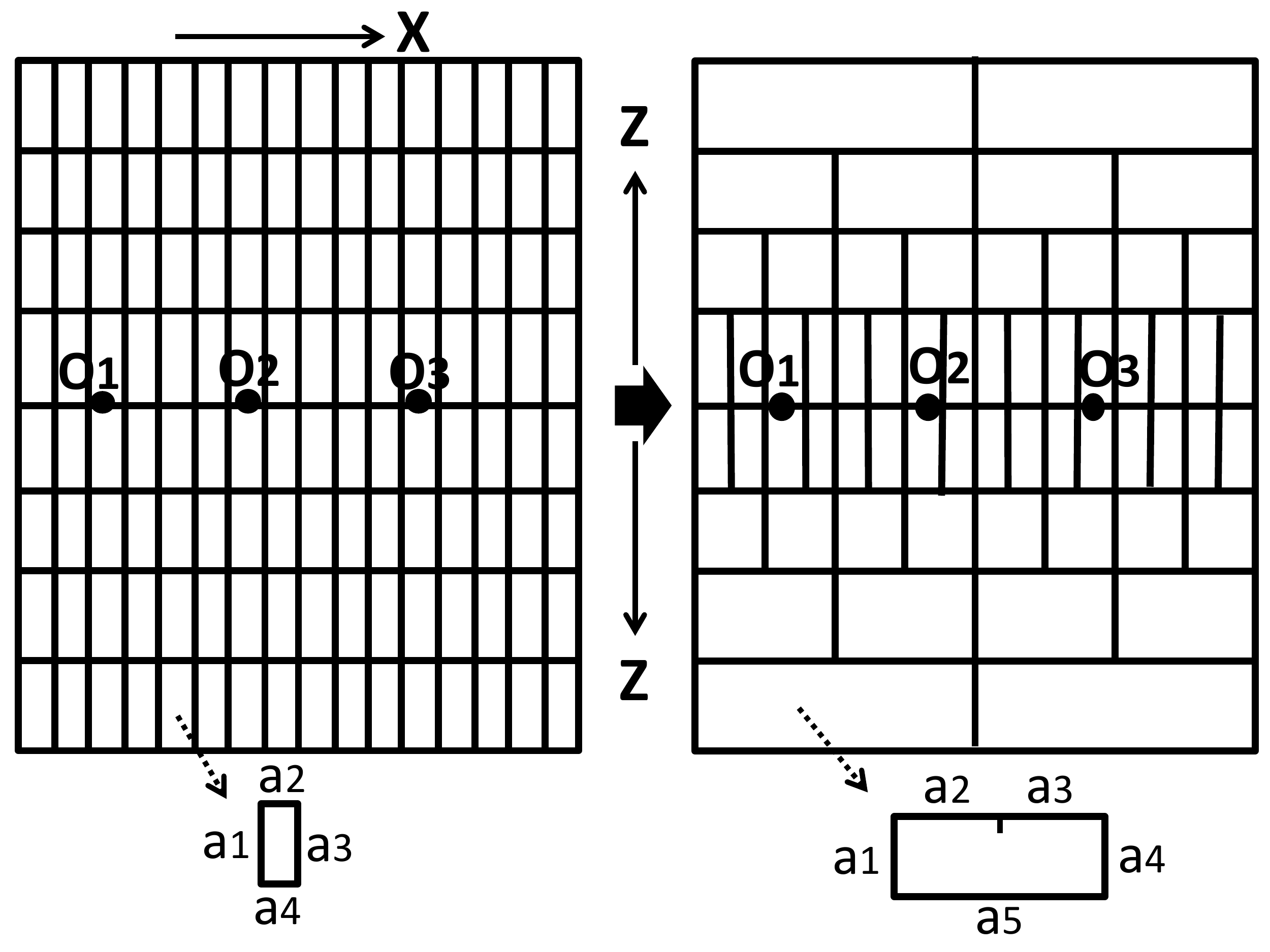}
 \caption{The sketch of the scale transformation and the correlation function. The left and right picture corresponds to the path-integral representation of correlation function before and after the optimization, respectively. A tensor is attached on the face of each of these cells. In the left picture there are tensor with 4 legs ($T_{a_1, a_2, a_3 a_4}$ ) attached on the faces of the cells. After the coarse graining procedure these tensors are replaced by some effective tensors ($T_{a_1, a_2, a_3, a_4, a_5}$) as  shown in the right picture. }
\label{fig:conformalTR}
\end{figure}

We would also like to point out the property of wave functional in CFTs (\ref{prptdtt}) follows from the well-known scaling property of primary operators in any CFTs (see e.g.\cite{Ga}):
\ba
&& \la O(x_1,z_1)O(x_2,z_2)\ddd O(x_n,z_n) \lb_{g=e^{2\phi}} \no
&& =\left(\prod^n_{i=1}e^{-\Delta_i\left(\phi(x_i,z_i)-\phi_0\right)}\right)\la O(x_1,z_1)O(x_2,z_2)\ddd O(x_n,z_n)\lb_{g=e^{2\phi_0}},  \label{primary}
\ea
where $\Delta_i=h_i+\bar{h}_i$ is the conformal dimension of each operator $O_i$. If we set
$z_1=z_2=\ddd= \epsilon$, then the condition (\ref{bclq}) tells us the overall factor in the RHS of (\ref{primary}) is one and thus the correlation functions at $z=\ep$ does not change under the
optimization. On the other hand, we can calculate this correlation function from the path-integration
\be
\la O_1(x_1)O_2(x_2)\ddd \lb=\frac{\int D\vp ~|\Psi_{g=e^{2\phi}}[\vp]|^2\cdot O_1O_2\ddd }{\int D\vp ~|\Psi_{g=e^{2\phi}}[\vp]|^2}. \label{idptha}
\ee
Here we regard the correlation functions as the path-integration over a space defined by a double copy of (\ref{metw}) with $O_i$ inserted on the center line. Refer to Fig.\ref{fig:conformalTR} for a sketch of this identity. The fact that the correlation functions is invariant under the optimization shows that the wave functional should be proportional to the original one as in (\ref{prptdtt}).
This provides another derivation of the claim (\ref{wsym}) for 2d CFTs.

\subsection{Path-integral Optimization of Perturbed 2D CFTs}

In this subsection we consider a deformation of a two dimensional CFT by a primary operator $O(x)$ with a position dependent coupling $\lambda(x,z)$ and a nontrivial metric. The action in flat space is
\be
S_{g_0,\lambda_0}[\vp]=S^{CFT}_{g_0}[\vp]+\ep^{2-\Delta}\lambda_0\int dx dz \s{g_0} ~ O(x,z), \label{actw}
\ee
where $\vp(x,z)$ represent all dynamical fields in the given CFT; $\Delta$ is the conformal
dimension of the primary scalar operator $O(x,z)$. The path-integration (\ref{phzo}) gives the
wave functional $\Psi_{g=e^{2\phi_0},\lambda_0}$ for the perturbed CFT vacuum.

Next we allow the metric and coupling $\lambda_0$ to depend on coordinates $x$ and $z$ in order to
optimize the path-integral. This way, the new deformed action looks like
\be
S_{g,\lambda}[\vp]=S^{CFT}_{g}[\vp]+\ep^{2-\Delta}\int dx dz \s{g}\lambda(x,z)O(x,z). \label{perto}
\ee
Next, we would like to focus on a special choice of $\lambda(x,z)$, written as $\lambda_\phi(x,z)$, such that the wave functional $\Psi_{g=e^{2\phi},\lambda_\phi}$, computed as in (\ref{psigl}), remains the same as the original one $\Psi_{g=e^{2\phi_0},\lambda_0}$
up to a normalization factor:
\ba
\Psi_{g=e^{2\phi},\lambda_\phi}[\vp(x)]=e^{N[e^{2\phi},\lambda_\phi]-N[e^{2\phi_0},\lambda_0]}\cdot
\Psi_{g=e^{2\phi_0},\lambda_0}[\vp(x)].  \label{wsym}
\ea
This condition fixes $\lambda(x,z)$ in terms of the function $\phi(x,z)$ and the initial condition $\lambda_0\equiv \lambda(x,\ep)$ (recall (\ref{pobc})). Note also
that we clearly have $\lambda_{\phi=\phi_0}=\lambda_0$ by definition. We will give a general argument how to choose $\lambda(x,z)$ which satisfies the relation (\ref{wsym}) in section \ref{more}.

The optimization can now be completed by minimizing the functional $N[e^{2\phi}, \lambda_\phi]$ with respect to $\phi(x,z)$. We will present explicit calculations of this procedure below in perturbation theory.

The claim (\ref{wsym}) is essentially equivalent to the following identity for any correlation functions at $z=\ep$ via the expression (\ref{idptha}):
\ba
\la O_1(x_1,\ep)O_2(x_2,\ep)\ddd O_n(x_n,\ep)\lb^{\lambda_\phi}_{g=e^{2\phi}}=
\la O_1(x_1,\ep)O_2(x_2,\ep)\ddd O_n(x_n,\ep)\lb^{\lambda_0}_{g=e^{2\phi_0}}.  \label{ecor}
\ea
Here we again defined the correlation function by the path-integration over a space defined by a double copy of (\ref{metw}) with $O_i$ inserted on the time slice $z=\ep$ (refer to Fig.\ref{fig:conformalTR} again). We can derive (\ref{ecor}) from (\ref{wsym}) by noticing that the correlation functions can be computed by taking derivatives with respect to sources $\lambda$, where we neglect contact terms which come from the derivatives of $N[e^{2\phi},\lambda_\phi]$ at $z=\ep$ (this will be explained in section \ref{ncont}).

\section{Free Field Examples}

In order to provide a support to our proposal, below we begin by analyzing our optimization procedure for the simplest models: free massive scalar field theory and free massive fermion theory in two dimensions.

\subsection{Massive Free Scalar}

Consider the action of a massive free scalar in a two dimensional space with a general metric \eqref{metw}
\be
S^{scalar}_{g,\lambda}=\frac{1}{2}\int dxdz \s{g}(g^{ab}\de_a \vp\de_b\vp)+\frac{1}{2}\ep^2\int dx dz \s{g}\lambda(x,z)\vp(x,z)^2,
\ee
where $\lambda(x,z)$ is the square of a position dependent mass. Let us choose $\lambda_\phi$ as follows
\be
\lambda (x,z)  = \lambda_\phi(x,z)=\lambda_0 e^{-2(\phi(x,z)-\phi_0)},  \label{massfs}
\ee
where we assume $\lambda_0$ is a constant. In this case, even when $\phi$ is non-trivial, we have
\be
S^{scalar}_{g,\lambda_\phi}=\frac{1}{2}\int dxdz \left((\de_x \vp)^2+(\de_z \vp)^2\right)+\frac{1}{2}\int dx dz \lambda_0\vp(x,z)^2,
\ee
which is just the action of a massive free scalar in the flat space.

To compute the wave functional we decompose the field into
\be
\vp(x,z)=\bar{\vp}(x,z)+\eta(x,z),
\ee
where $\bar{\vp}$ is the classical solution to equation of motion with the boundary condition
$\bar{\vp}(x,z=0)=\vp(x)$. It is explicitly written using Fourier transform as follows
\ba
\bar{\vp}(x,z)=\int^\infty_{-\infty}dk\vp(k)e^{ikx-\s{k^2+\lambda_0}(z-\ep)}.
\ea
The function $\eta(x,z)$ (with vanishing boundary condition at $z=\epsilon$) describes quantum fluctuations around the classical solution.
Then we get the wave functional
\ba
&& \Psi_{g=e^{2\phi},\lambda_\phi}[\vp(x)] \no
&& =e^{-S^{scalar}_{g_0,\lambda_0}[\vp]}\times \int \prod_{x,z}[D\eta(x,z)]e^{-\frac{1}{2}\int dxdz \left[(\de\eta)^2+\lambda_0\eta^2\right]}\cdot \prod_{x}\delta \left[\eta(x,z=0)\right] ,
\ea
where $ S^{scalar}_{g_0,\lambda_0}[{\vp}]$ gives the classical contribution (we performed the Fourier transformation)
\be
S^{scalar}_{g_0,\lambda_0}[\vp]=\pi\int dk\s{k^2+\lambda_0}\ \vp(-k)\vp(k).
\ee

It is obvious that this classical part does not depend on the Weyl factor $\phi(x,z)$. The other factor is the partition function of $\eta$ with the Dirichlet boundary condition $\eta(x,z=0)=0$ and does not depend on the field $\vp(x)$. The latter gives an overall normalization factor which depends on $\phi$ and $\lambda_0$:
\ba
e^{N[e^{2\phi},\lambda_\phi]}= \int \prod_{x,z}[D\eta(x,z)]_{g=e^{2\phi}}~e^{-\frac{1}{2}\int dxdz \left[(\de\eta)^2+\lambda_0\eta^2\right]}\Biggr |_{\eta(x,z=0)=0}.
\ea
Note that the above path-integration over $\eta$ non-trivially depends on $\phi$ because the path-integration is performed with an integration measure with a UV cut off which is specified by the metric $g=e^{2\phi}$.
If we set $\lambda_0=0$, then $N[e^{2\phi},0]$ agrees with the Liouville action $S_L[\phi]$ as the theory gets conformally invariant. For general values of $\lambda_0$, the full evaluation of $N[e^{2\phi},\lambda_\phi]$ is not straightforward. We will discuss results to the first order perturbation of the mass square $m_0^2=\lambda_0$ in the appendix \ref{Mabuchi}, employing the calculations in \cite{FKZ}. In this analysis, however, we find non-standard results with logarithmic terms because the operator $\vp^2$ in the massless scalar CFT is not a primary operator in two dimensions.

\subsection{Massive Free Fermions}
The derivation for massive free fermion is analogous to the scalar case. We start with the two dimensional action in metric \eqref{metw}
\be
S^{fermion}_{g,m}=\int dxdz\sqrt{g}\bar{\psi}\gamma^\mu\nabla_\mu\psi-\ep\int dxdz\sqrt{g}\,m(x,z)\bar{\psi}\psi.
\ee
Then, in the Weyl rescaled metric \eqref{metw}, fermion fields are transformed as
\be
\psi(z,x)\to e^{-\frac{1}{2}\phi(z,x)}\psi(z,x),\qquad \bar{\psi}(z,x)\to e^{-\frac{1}{2}\phi(z,x)}\bar{\psi}(z,x),
\ee
however, the covariant derivative with spin connection renders the kinetic term Weyl invariant (free fermion CFT).

On the other hand, we fine-tune the space dependent mass to
\be
m_\phi(x,z)=m_0 e^{-(\phi(x,z)-\phi_0)},
\ee
and cancel the additional Weyl factor from the metric determinant. This way, the full classical action in our Weyl-rescaled metric becomes the free massive fermion in flat space
\be
S^{fermion}_{g,m_\phi}=\int dxdz\bar{\psi}\left(\gamma^\mu\partial_\mu-m_0\right)\psi.
\ee
Now, when computing the wave functional, we have to be a bit more careful than for bosons. Namely, the equations of motion for fermions read
\be
\gamma^\mu\partial_\mu\psi-m\psi=0,\qquad \partial_\mu\bar{\psi}\gamma^\mu+m\bar{\psi}=0.\label{eomF}
\ee
Then, in the computation of the wave functional, we split the field into the classical part with a given boundary condition and a quantum part that vanishes at the boundary
\be
\bar{\psi}=\bar{\psi}_c+\bar{\eta},\qquad \psi=\psi_c+\eta.
\ee
If we now impose equations \eqref{eomF} for $\bar{\psi}_c$ and $\psi_c$ , the classical action vanishes. To cure this problem we follow Wilczek and Larsen \cite{Larsen:1994yt} that discussed a similar issue for the mass less case. Their main point is that, for uniqueness of the classical fermion fields with fixed boundary conditions, it is enough to impose the Klein-Gordon equation (and not Dirac) obtained by application of the conjugate operators to \eqref{eomF}
\be
\left(\partial^\mu\partial_\mu-m^2_0\right)\psi_c=\left(\partial^\mu\partial_\mu-m^2_0\right)\bar{\psi}_c=0.
\ee
From there, with analogy to the free massive boson, we perform the path integral over $\bar{\eta}$ and $\eta$ that leaves us with the ratio of the wave functionals
\be 
\frac{\Psi[\bar{\psi},\psi]_{g=e^{2\phi}}}{\Psi[\bar{\psi},\psi]_{g_0=e^{2\phi_0}}}\equiv e^{N[g,m_\phi]-N[g_0,m_0]}=\frac{det\left[\gamma^\mu\partial_\mu-m_\phi\right]}{det\left[\gamma^\mu\partial_\mu-m_0\right]}.
\ee
Clearly, as for bosons, this ratio will depend on the field $\phi$ that we use in the computation of the determinant in the numerator (UV regulator specified by the rescaled metric).\\
Note that for fermions the mass term is a primary operator so this example satisfies the requirements of our perturbative computations. It would also be very interesting to generalize Mabuchi action approach \cite{FKZ} to fermions.

\section{Optimizing Relevant Perturbations in 2D CFTs}\label{more}

In this section we would like to explicitly analyze the path-integral optimization procedure in general
two dimensional CFTs with relevant perturbations. We will analyze leading contributions in perturbation theory in a position dependent coupling. For earlier arguments on RG flows with the position dependent couplings (local RG flows) refer to e.g. \cite{Os,FK,Nk}.

\subsection{Wilsonian RG Flow}

As we explained, the key idea of the path-integral optimization was that we keep the same quantum state even if we Weyl-rescale the metric by $e^{2\phi}$, by locally changing the coupling constants.
Let us interpret this in terms of RG flows. In RG flows, we change the energy scale without changing the physics. In particular, we take $\Lambda$ to be the cut off scale in the Wilsonian sense and consider the coupling constant at this scale  $\lambda(\Lambda)$. The theory does not change if $\lambda(\Lambda)$ satisfies the RG equation
\be
\Lambda \frac{d \lambda(\Lambda)}{d\Lambda}=\beta\left(\lambda(\Lambda)\right),
\ee
where $\beta$ is the beta function. The partition function $Z$ and correlation function satisfy the Callan-Symanzik equation
\ba
\left(\Lambda\frac{d}{d\Lambda}+\beta(\lambda)\frac{d}{d\lambda}+\xi(\lambda)\right)Z[\Lambda,\lambda]=0,
\label{CSeq}
\ea
where $\xi(\lambda)$ represents extra contributions which break the conformal symmetry, such as the conformal anomaly.

\subsection{Local RG Flow}

In our setup with the metric (\ref{metw}), our rule was that the size of a single lattice site is the unit area in this metric. Therefore if we write the size of lattice in the coordinate $x$ and $z$, we have
\be
\Delta x=\Delta z=e^{-\phi}.\label{lattice}
\ee
 Remember that the original fine-grained lattice corresponds to $e^{\phi}=1/\ep\equiv e^{\phi_0}$ or  equivalently $\Delta x=\Delta z=\ep$. The regularization (\ref{lattice}) can be identified with the cut off scale $\Lambda$ in the Wilsonian RG flow via the relation
\be
\Lambda=e^\phi.
\ee
 The large/small $\phi$ corresponds to the UV/IR limit. Therefore, in order not to change the theory, we need to take $\lambda=\lambda_\phi$ such that
\be
\frac{d\lambda_\phi}{d\phi}=\beta(\lambda_\phi). \label{rgb}
\ee
The beta function for the operator $O(x)$ with the dimension $\Delta$ is given in a standard way
\ba
\beta(\lambda)=(\Delta-2)\lambda+O(\lambda^2).\label{betew}
\ea

Now, for our path-integral optimization, we need to introduce a position dependent cut off as we already explained before. Thus we need to regard $\Lambda=e^\phi$ as a function of coordinates
$(x,z)$. Accordingly, we need to consider a local version of the beta function equations (\ref{rgb}), (\ref{betew}) which look like
\ba
 \f{\delta \lambda_{\phi}(x,z)}{\delta \phi(x',z')}
=\beta\left[\lambda_\phi\right]^{(x,z)}_{(x',z')},\label{betgh}
\ea
where
\ba
\beta\left[\lambda\right]^{(x,z)}_{(x',z')}=(\Delta-2)\lambda(x,z)\cdot \delta(x-x')\delta(z-z')
+O(\lambda^2).
\ea
Even though the higher order terms $O(\lambda^2)$ are expected to be non-local in general,
we can neglect such higher order contributions in our leading order analysis below.
By solving (\ref{betgh}) with the boundary condition
$\lambda_{\phi}|_{\phi=\phi_0}=\lambda_0$, we obtain the form of
$\lambda_\phi$ to the leading order of $\lambda$ expansion as follows
\be
\lambda_\phi= e^{(\Delta-2)(\phi-\phi_0)}\cdot \lambda_0 +O(\lambda_0^2).  \label{lowl}
\ee

Similarly, the Callan-Symanzik equation (\ref{CSeq}) is locally written as
\ba
\left(\frac{\delta}{\delta\phi(x,z)}+\int dx'dz'\beta\left[\lambda\right]^{(x',z')}_{(x,z)}\cdot\frac{\delta}
{\delta\lambda(x',z')}+\xi[\lambda]_{(x,z)}\right)Z[\phi,\lambda]=0.
\label{CSPeq}
\ea
In particular, when the relevant perturbation is turned off, $\lambda=0$, we have $\xi(x,z)=\frac{c}{12\pi}(\de_x^2+\de_z^2)\phi(x,z)$ and the solution is given in terms of the Liouville action as $Z[\phi]=e^{S_L[\phi]}$ as expected to be true for any two dimensional CFTs.

More generally, if we focus on the partition function $Z[\phi,\lambda_\phi]$, which only depends on $\phi$ and
$\lambda_0$, we can rewrite (\ref{CSPeq}) into
\ba
\left(\frac{\delta}{\delta\phi(x,z)}+\xi[\lambda_\phi]_{(x,z)}\right)Z[\phi,\lambda_\phi]=0,
\label{CSPeqq}
\ea
where $\frac{\delta}{\delta\phi(x,z)}$ should be interpreted as the total derivative as opposed to
the partial derivative in (\ref{CSPeq}). This is formally solved as
\ba
\frac{Z[\phi,\lambda_\phi]}{Z[\phi_0,\lambda_0]}=\exp\left[-\int dx dz \int^{\phi(x,z)}_{\phi_0} \delta \phi(x,z)\xi[\lambda_\phi]_{(x,z)}\right]. \label{partfac}
\ea

Now if we consider the path-integral for the vacuum wave functional (\ref{psigl}) with the boundary condition (\ref{pobc}), the above property (\ref{partfac}) leads to the relation (\ref{wsym}) for the choice of $\lambda_\phi$ given by (\ref{betgh}) and (\ref{lowl}).

\subsection{Evaluation of the Normalization $N[e^{2\phi},\lambda_{\phi}]$}\label{ncont}

In this subsection we use perturbation theory to calculate the normalization functional $N[e^{2\phi},\lambda_\phi]$ in (\ref{wsym}), which measures the path-integral complexity. For simplicity we focus on the case where the coupling $\lambda_0$ is constant. In this case we can assume that $\lambda_\phi$ only depends on $z$.

In perturbation expansion in $\lambda_0$ the normalization can be written in terms of correlators in the CFT defined on the upper half plane with a boundary at $z = \ep$.
\ba
 N[e^{2\phi},\lambda_\phi]= S_L[\phi] + N_{1pt}[e^{2\phi},\lambda_\phi] + N_{2pt}[e^{2\phi},\lambda_\phi] + O(\lambda_\phi^3)
\label{pta1}
\ea
where the first order contribution is
\ba
N_{1pt}[e^{2\phi},\lambda_\phi]=-\ep^{2-\Delta}\int dx dz e^{2\phi(x,z)}\lambda_{\phi}(z)\la O(x,z)\lb
\label{pta2}
\ea
while the second order contribution is
\ba
 N_{2pt}[e^{2\phi},\lambda_\phi]  &=& \frac{1}{2}\ep^{4-2\Delta}\left(\prod^2_{i=1}\int dx_i dz_i\s{g} \right) \lambda_\phi(z_1)\lambda_\phi(z_2) \no 
&&\times \Bigl(\la O(x_1,z_1)O(x_2,z_2)\lb-\la O(x_1,z_1)\lb\la O(x_2,z_2)\lb\Bigr).  \label{ptb}
\ea
Notice that at this order, $\lambda_{\phi}$ is proportional to $\lambda_0$ as in (\ref{lowl}).

Consider first $N_{1pt}[e^{2\phi},\lambda_\phi]$. This is non-vanishing because the CFT correlator has to be evaluated on the upper half plane whose boundary is at $z=\ep$. The full answer depends on the boundary condition.
On the other hand, if we assume the relation (\ref{wsym}), the functional $N[e^{2\phi},\lambda_\phi]$ should not depend on the boundary condition \footnote{Indeed explicit calculations in free scalar field theory show that the divergent contributions are independent of the boundary conditions.}. Therefore, to analyze this boundary contribution, we can take the simplest boundary condition e.g. a conformal boundary condition for the basic field. Then, one point function with our appropriately prescribed cut off is given by
\ba
\la O(x,z)\lb=b_0\cdot\frac{e^{-\Delta\phi(x,z)}}{(z^2+e^{-2\phi(x,z)})^{\Delta/2}}.
\ea
for some constant $b_0$. This leads to
\ba \label{intgr}
N_{1pt}[e^{2\phi},\lambda_\phi]&=&-\ep^{2-\Delta}\int dx dz~ e^{2\phi(x,z)}\lambda_{\phi}(z)\la O(x,z)\lb \no
&\simeq &-b_0\int dx \int^\infty_\ep dz\frac{\lambda_0}{(z^2+e^{-2\phi(x,z)})^{\Delta/2}}\no
&\simeq & -b\lambda_0\int dx~ e^{(\Delta-1)\phi(x,\ep)} \no
&=& -b\lambda_0\ep^{1-\Delta}\int dx , \label{onepy}
\ea
where $b$ is a constant proportional to $b_0$ up to a $O(1)$ positive constant, which depends on the detail of the UV regularization; we again ignored $O(\lambda_0^2)$ terms in (\ref{lowl}) because
the integral is localized at $z=\ep$, where we have $\lambda_\phi=\lambda_0$.  Also we have assumed  $\Delta > 1$ while performing the integral in (\ref{intgr}) to avoid IR divergence.
Notice that in principle, the sign of $b$ depends on the choice of the relevant perturbation. If we consider a mass perturbation of scalar field theories, it is obvious that we have $b>0$ (and $b_0>0$). If the operator $O$ has a Z$_2$ symmetry $\lambda_0\lr -\lambda_0$, then a Z$_2$ symmetric boundary condition leads to $b=b_0=0$, for example in the case of free fermion theories.

Note that this contribution does not depend on the metric $\phi$. As we will see soon the one point function does not contribute to our optimization procedure, but it gives a constant divergent term for the total complexity (see below), assuming $1<\Delta<2$.

Let us now consider the $O(\lambda_0^2)$ contribution, $N_{2pt}[e^{2\phi},\lambda_\phi]$. We will first evaluate this ignoring the boundary at $z=\ep$, denoting this by $N_{2pt}^{plane}[e^{2\phi},\lambda_\phi]$. We will then consider the effect of the boundary.
If we use the formula (\ref{primary}) for CFTs (for that we need to take the CFT point $\lambda=0$)
then we immediately find that if the integrals over $x_{1,2}$ and $z_{1,2}$ converged, then we would find that this contribution to $N[e^{2\phi},\lambda_\phi]-N[e^{2\phi_0},\lambda_0]=S_L[\phi]-S_L[0]$, i.e. coincide with the conformally invariant case. However, there are actually UV divergences when $x_1,z_1$ and $x_2,z_2$ get closer in the integrals, which we have to regulate.\\
For this, we introduce a position dependent UV cut off such that we define the full plane two-point function of the operators on the Weyl rescaled background \eqref{metw} to be
\be
\la O(x_1,z_1)O(x_2,z_2)\lb_{\phi}^{plane}= \frac{e^{-\Delta \phi(x_1,z_1)-\Delta\phi(x_2,z_2)}}
{\left(|x_1-x_2|^2+|z_1-z_2|^2+e^{-\phi(x_1,z_1)-\phi(x_2,z_2)}\right)^{\Delta}},
\label{regp}
\ee
where $\Delta=h+\bar{h}$ is the conformal dimension of the primary field $O$.
Note that at the UV point $e^{\phi}=1/\ep$ we have the fine-grained regularization $\sim \left(|x_1-x_2|^2+|z_1-z_2|^2+\ep^2\right)^{-\Delta}$ as usual. For more general values of $\phi$, the above choice follows from the position dependent cut off (\ref{lattice}).

This way, by performing the regularized integrals, the quadratic term in (\ref{ptb}) is evaluated as follows \footnote{
We used the formula $\int^\infty_0 \frac{rdr}{(r^2+a^2)^\Delta}=\frac{1}{2(\Delta-1)} a^{2-2\Delta}$, when $\Delta > 1$.}:
\ba
N_{2pt}^{plane} & = & \frac{1}{2}\ep^{4-2\Delta}\int dx_1 dz_1 e^{2\phi(x_1,z_1)}\int dx_2 dz_2 e^{2\phi(x_2,z_2)} \lambda_\phi(z_1)\lambda_\phi(z_2)\la O(x_1,z_1)O(x_2,z_2)\lb \no
&& \simeq \frac{1}{2}\int dx_1dz_1\int dx_2 dz_2 \frac{\lambda_0^2}
{\left(|x_1-x_2|^2+|z_1-z_2|^2+e^{-\phi(x_1,z_1)-\phi(x_2,z_2)}\right)^{\Delta}} \no
&&\simeq \frac{\lambda_0^2}{4(\Delta-1)}\int dx dz~ e^{(2\Delta-2)\phi(x)}.
   \label{ndla}
\ea
In the above computation we kept only terms which do not vanish in the continuum limit
$\ep\to 0$. We neglect the derivative terms because they accompany positive power of
$\ep\sim e^{-\phi}$ for example,
$\lambda_0^2e^{(2\Delta-4)\phi}(\de_z\phi)^2\sim \ep^{4-2\Delta}$.
Also note that we need to impose $1<\Delta<2$ so that the term (\ref{ndla}) satisfies this non-vanishing condition in the $\ep\to 0$ limit. This range of conformal dimension corresponds to the standard quantization ($\Delta_+$ quantization) in the AdS/CFT \cite{KlWi}. It will be an interesting future problem to work out how the alternative quantization ($\Delta_-$ quantization) can be realized in our formalism.

The expression for $N_{2pt}[e^{2\phi}, \lambda_\phi]$ in (\ref{ndla}) needs modification due to the presence of the boundary at $z=\epsilon$. Pretty much like the one point function, we expect that the divergent terms do not depend on the boundary values of the basic fields of the CFT and we can use conformal boundary conditions. This is because the divergence arises only either when two bulk points get closer
or when at least one of them get closer to the boundary. The latter contributions have been already
taken into account in the one point function contribution $N_{1pt}[e^{2\phi}, \lambda_\phi]$ (\ref{onepy}).
We also expect that the answer is independent of the Liouville mode.

In this way, we find that $N_{2pt}[e^{2\phi},\lambda]$ is given by $N^{plane}_{2pt}[e^{2\phi},\lambda]$ i.e.
(\ref{ndla}). Clearly, this shows that the normalization factor $e^{N[e^{2\phi},\lambda_\phi]}$ depends on the coupling $\lambda$ non-trivially. Nevertheless, if we take functional derivatives of $N[e^{2\phi},\lambda_\phi]$ with respect to $\lambda_\phi(x,z=\ep)=\lambda_0(x,z=\ep)$, this extra contributions from $N[e^{2\phi},\lambda_\phi]$ are all delta
functional terms $\delta(x_1-x_2)\cdots$, which we can neglect. This supports the derivation of (\ref{ecor}) as well as our claim (\ref{wsym}).

In principle, we can proceed with perturbative expansion to higher orders and, if we keep only terms which survive the UV limit $\ep\to 0$, we expect the following structural form of the normalization functional
\ba
N[e^{2\phi},\lambda_\phi]=S_{L}[\phi]+N_{1pt}[e^{2\phi},\lambda_\phi]+\int dx dz \sum_{n=2}^\infty h_n \cdot(\lambda_0)^{n}e^{\left(2+(\Delta-2)n\right)\phi},  \label{generaln}
\ea
where $h_n$ are numerical coefficients related to the $n$ point function contribution.

\subsection{Complexity Functional}

Based on  \cite{Caputa:2017urj}, we regard the normalization functional $N[e^{2\phi},\lambda_\phi]$ as a measure of complexity of each wave functional. As stated in (\ref{pcdef}), the minimum of this functional with respect to the variations of the metric gives a measure of complexity of the given quantum state, called the path-integral complexity.

As we have already stressed, we assume that the dimension of $O$ is in the range $1<\Delta<2$.
Then, our perturbative results up to the quadratic order lead to the following complexity functional \footnote{it is convenient to redefine $\lambda_0$ as $\frac{1}{4(\Delta-1)}\lambda^2_0\to \frac{c}{24\pi}\lambda^2_0$}
\ba
&& N[e^{2\phi},\lambda_\phi]  \simeq S_L[\phi]+N_{2pt}[e^{2\phi},\lambda_\phi]+N_{1pt}[e^{2\phi},\lambda_\phi] \no
&&=\frac{c}{24\pi}\int dxdz \left[(\de
\phi)^2+e^{2\phi}+\lambda_0^2 e^{(2\Delta-2)\phi} \right]-b\ep^{1-\Delta}\lambda_0\int dx, \label{fffgq}
\ea
which only depends on $\phi$ and $\lambda_0$.

The path-integral complexity $C[\lambda_0]$, as given in (\ref{pcdef}), is obtained by minimizing
$N[e^{2\phi},\lambda_\phi]$ by varying the function
$\phi(z)$, imposing the boundary condition $\phi(z=\ep,x)=\phi_0$ and the relation (\ref{lowl}).
The value of the coefficent $b$ (and $b_0$) is expected to be $O(c)$ in general as in the holographic dual of BCFT \cite{AdSBCFT}.

Let us now consider the solution of the minimization procedure that gives the condition
\ba
\frac{\delta N[e^{2\phi},\lambda_\phi]}{\delta\phi}=0.
\ea
Given our path integral general complexity functional \eqref{fffgq}, we can now find the perturbative correction around the $\lambda_0=0$ solution (\ref{hypas}) as follows
\ba
e^{\phi(z)}=z^{-1}\left(1-\frac{\lambda^2_0}{2(5-2\Delta)}z^{-2\Delta+4}+\ddd\right).
\label{phiex}
\ea
This result agrees with our intuitive expectations. Namely, the presence of a relevant perturbation
reduces the degrees of freedom in the IR region and we can coarse-grain the path-integral more in the IR region. Indeed, the function $e^{\phi(z)}$ of (\ref{phiex}) is reduced in the IR region.

Next, from the optimized metric, we we can estimate a perturbative correction to the path integral complexity from the relevant perturbation. Notice that we can expand
\ba
&& (\de\phi)^2+e^{2\phi}\simeq 2z^{-2}+\frac{3-2\Delta}{5-2\Delta}\lambda_0^2 z^{2-2\Delta}+\ddd,\no
&&\lambda_0^2 e^{(2\Delta-2)\phi} \simeq \lambda_0^2 z^{2-2\Delta}+\ddd.
\ea
Therefore the change of the path-integral complexity is finally evaluated as
\ba
&& C[\lambda_0]-C[0] \simeq L\lambda^2_0
\int^\infty_\ep dz \frac{4(2-\Delta)}{5-2\Delta}z^{2-2\Delta}-bL\lambda_0\ep^{1-\Delta},
\label{cpxfn}
\ea
where $L=\int dx$  is the total length of the spatial direction.

When $\frac{3}{2}<\Delta<2$, the first contribution in (\ref{cpxfn}) gets UV divergent as
\ba
 C[\lambda_0]-C[0]\simeq \frac{4(2-\Delta)}
{(5-2\Delta)(2\Delta-3)} L\lambda^2_0\ep^{3-2\Delta}-bL\lambda_0\ep^{1-\Delta}.  \label{vvv}
\ea
It is worth noting that the bulk contribution (i.e. the first term in (\ref{vvv}) on the RHS) is always positive. The reason for this positive bulk contribution is obvious because the additional perturbative term is quadratic with the positive coefficient. Also note that, in particular when $\Delta=3/2$, we expect a logarithmic contribution $\sim -L\lambda^2_0\log\ep$.

On the other hand, if  $1<\Delta<\frac{3}{2}$, we need to remember that the expansion (\ref{phiex})
breaks down when $z$ gets larger such that $\lambda_0 z^{-\Delta+2}\sim 1$. Thus the first term in (\ref{cpxfn}) is estimated to the following finite contribution:
\ba
\sim\int^{(\lambda_0)^{\frac{1}{\Delta-2}}}_\ep dz \frac{4(2-\Delta)}{5-2\Delta}z^{2-2\Delta}
= \frac{4(2-\Delta)}
{(5-2\Delta)(3-2\Delta)}(\lambda_0)^{\frac{3-2\Delta}{\Delta-2}},
\ea
which leads to the behavior
\ba
 C[\lambda_0]-C[0]\sim  L(\lambda_0)^{\frac{1}{2-\Delta}}-bL\lambda_0\ep^{1-\Delta}. \label{abgf}
\ea

In both the cases the leading contribution to $(C[\lambda_0]-C[0])$ comes from the one-point function due to the presence of the boundary (i.e. the second term in (\ref{cpxfn}) on the RHS), however the sign of this contribution depends on the choice of the relevant perturbation.

\section{Comparison with AdS/CFT}

Finally we would like to explore a possible connection between our path-integral optimization result for the relevant perturbation and known results in the AdS/CFT correspondence. For that, we consider the setup of Einstein gravity coupled to a single massive scalar field $\Phi$ with the mass $M$ in $d+1$ dimensions
\ba
I_{gravity}=\frac{1}{16\pi G_N}\int d^{d+1}x\s{-G}\left[R-\frac{1}{2}(\de\Phi)^2+\frac{d(d-1)}{R_{AdS}^2}-\frac{1}{2}M^2\Phi^2\right].
\ea
We will set $R_{AdS}=1$ below.  The conformal dimension of the operator $O$ dual to $\Phi$ is given by the holographic dictionary $\Delta=\frac{d}{2}+\s{\frac{d^2}{4}+M^2}$. When we add the external perturbations given by (\ref{perto}), accordingly to the standard bulk to boundary relation, the solution of the scalar field $\Phi$ looks like
\ba
\Phi(z,x)=z^{d-\Delta}\lambda_0(x)+z^{\Delta}\la O(x)\lb+\ddd.
\ea
This motivates us to identify the bulk scalar in AdS as our running coupling constant (\ref{lowl}). Namely, to the leading order
\be
\Phi(z,x)\sim \ep^{d-\Delta}\lambda_\phi(z,x)=\lambda_0(x) e^{(\Delta-d)\phi}+O(\lambda_0^2).
\ee

Now let us focus on the AdS$_3/$CFT$_2$ (i.e. $d=2$) and aim to solve the Einstein equations for the back-reacted metric from the scalar field. We can refer to \cite{HMS} for the perturbative solution to the Einstein equation. We assume $\lambda(x)$ is a constant $\lambda_0$. It takes the form\footnote{Here, we neglect the expansions from the normalizable mode $f(y)=\ddd+y^2(b_1+...)$.}
\ba
&& ds^2=\frac{1}{y^2}\left(dy^2+f(y)(-dt^2+dx^2)\right), \no
&& f(y)=1-\frac{\lambda^2_0}{4}y^{4-2\Delta}+\sum_{k=1}^\infty a_k \left(\lambda_0 y^{2-\Delta}\right)^{k+2},  \label{metexp}
\ea
where $a_k$ are numerical coefficients, which are computable.

If we take the time slice $t=0$ and redefine the coordinate $y$ into $z=z(y)$ so that  the metric takes the conformal gauge form \eqref{metw}, then we find
\ba
&& e^{\phi(z)}\simeq z^{-1}\left(1-\frac{2-\Delta}{ 4(5-2\Delta)}\lambda^2_0z^{4-2\Delta}\right),
\ea
where we used the perturbative map between $z$ and $y$
\ba
 z(y)\simeq y\left(1+\frac{\lambda^2_0}{ 8(5-2\Delta)}y^{4-2\Delta}\right). \label{ert}
\ea
This metric agrees with (\ref{phiex}) up to a numerical $O(1)$ factor, which depends on the details of UV regularization and normalization of the operator $O$. It is also intriguing to note that the form of $f(z)$ (\ref{metexp}) w.r.t $\lambda_0$ perturbation agrees with the general form (\ref{generaln}).

Last but not the least, this match supports our proposal that the space obtained from the path-integral optimization (via the minimization of the path-integral complexity) describes the canonical time slice in the gravity dual.

\section{Conclusions and Discussions}

In this work, we explored the path-integral optimization and the path-integral complexity for two dimensional CFTs with relevant perturbations. We pointed out the invariance of
correlation functions (\ref{ecor}) under the optimization is essentially equivalent to the basic condition (\ref{wsym}) of optimization for wave functionals. Since the former follows from the conformal invariance, we could derive the latter, which is the basic assumption of path-integral optimization, for CFTs. We gave an argument based on local RG flows that supports our condition (\ref{wsym}) for two dimensional CFTs with a relevant perturbation 
by a primary operator $O$. This allowed us to formulate the path-integral optimization and complexity for two dimensional CFTs with relevant perturbations. We explicitly calculated the leading perturbative contribution to path-integral complexity as well as the optimized metric.

Our path-integral complexity has two contributions in the leading order of the relevant perturbation
$\lambda_0$: the bulk term $\Delta C_{bulk}$ and the boundary term $\Delta C_{bdy}$. The latter comes from the one point function of the CFT in the presence of a boundary. This leads to the divergent term $\Delta C_{bdy}=-bL\ep^{1-\Delta}$, where $L$ is the spatial length of our two dimensional spacetime. The sign of $b$ depends on the choice of $O$. The sign of $b$ depends on the field theory we consider. For example, $b$ is positive for the mass deformation $\int \lambda_0\vp^2$ in the free scalar field theory. In the free fermion theory, we expect $b=0$ due to the chiral symmetry.

The bulk term $\Delta C_{bulk}$ is quadratic in the leading order of $\lambda_0$.
When the conformal dimension $\Delta$ of the primary operator $O$ is in the range $\frac{3}{2}<\Delta<2$, we find the positively divergent contribution $\Delta C_{bulk}\propto L\ep^{3-2\Delta}\lambda_0^2$ as in (\ref{vvv}).
When $\Delta=\frac{3}{2}$, this becomes a logarithmic divergence $\Delta C_{bulk}\propto -L\lambda_0^2 \log\ep$. When $1<\Delta<\frac{3}{2}$, we only find a finite contribution $\Delta C_{bulk}\propto (\lambda_0)^{\frac{1}{2-\Delta}}$ to the path-integral complexity as in (\ref{abgf}). By comparing this with the AdS/CFT analysis, we find a matching between our optimized metric and the back-reacted metric of time slice on an asymptotically AdS with a scalar field perturbation.

It is also intriguing to compare our results with holographic complexity. Firstly, the holographic complexity equals volume conjecture \cite{SUR,Susskind}, which is equivalent to the holographic quantum information metric \cite{InfoM}, corresponds to the choice of different complexity functional $C_V[\phi,\lambda_0]=\int dxdz e^{2\phi}$ in our formulation, where the optimization is done by minimizing the action (\ref{fffgq}). It is obvious in this calculation that the complexity decreases under relevant perturbations because the metric gets reduced (see also \cite{Carmi:2017ezk,Flory}).
The contribution proportional to $V\lambda_0^2\ep^{d+1-2\Delta}$ was found in \cite{Alish,Rath} for
AdS$_{d+1}/$CFT$_{d}$ setup, whose form coincides with our $\Delta C_{bulk}$.

Secondly, the holographic complexity equals action conjecture \cite{BrownSusskind1,Lehner:2016vdi} looks analogous to our formulation presented in this paper based on the action (\ref{fffgq}) at first sight. For example, the perturbative contribution of the form $V\lambda_0^2\ep^{d+1-2\Delta}$ was observed in a specific example in \cite{Moosa}. However, we find two open issues. In our formulation, the leading UV divergence follows volume law $\sim \ep^{-(d-1)}$ as it follows from the Liouville action, while the holographic action proposal leads to the logarithmic enhancement $\sim \ep^{-(d-1)}\log\ep$ \cite{CMR}. Another point is that for the relevant perturbation $\int \lambda_0 O$, we have a boundary contribution $\Delta C_{bdy}$ which is linear with respect to $\lambda_0$, while we do not expect such a term in the holographic complexity as the bulk supergravity action does not include any liner term. This term might be related to contributions from null boundaries.

It will be also interesting to estimate the circuit complexity for perturbed conformal field theories  and compare with our answers \cite{AB}. Refer to \cite{Chapman:2017rqy,JeMy,Molina-Vilaplana:2018sfn,KKS,HaMy} for computations in massive free field theories. To compare with results in the actual AdS/CFT, we need to study special features of holographic CFTs, which have a large central charge and are strongly coupled. We expect that they will appear at higher orders of perturbation theory in our path-integral optimization approach. Finally, it remains to be a very important direction to work out how to estimate the path-integral complexity for time-dependent states.

\section*{ Acknowledgements}

We are grateful to Mohsen Alishahiha, Jose Barbon, Ling-Yan Hung, Alexander Jahn, Jose J. Fernandez-Melgarejo, Rob Myers, Yu Nakayama, Aninda Sinha, Michael Smolkin, Kento Watanabe for useful discussions and correspondences. AB is supported by JSPS fellowship. AB and TT are supported by JSPS Grant-in-Aid for JSPS fellowship 17F17023. PC and TT are supported by the Simons Foundation through the ``It from Qubit'' collaboration. PC is supported by the JSPS starting grant KAKENHI 17H06787 .  NK and TT are supported by JSPS Grant-in-Aid for Scientific Research (A) No.16H02182. TT is also partially supported by World Premier International Research Center Initiative (WPI Initiative) from the Japan Ministry of Education, Culture, Sports, Science and Technology (MEXT). The work of SRD is partially supported by National Science Foundation grant NSF-PHY/1521045. SRD thanks Yukawa Institute for Theoretical Physics for hospitality during the initial stages of this project.

\appendix
\section{Analysis of Massive Free Scalar} \label{Mabuchi}

Here we present a perturbative analysis of path-integral optimization for a free scalar in two dimensions by employing the results in \cite{FKZ}.
We regard the mass term as a relevant perturbation $\lambda=m^2$ and calculate its contribution to the normalization factor $N[e^{2\phi},\lambda_\phi]$. Note that since the mass term $\int m^2\phi^2$ is not a primary operator in two dimensions, our results below do not match with our analysis in section 5.
Indeed the leading contribution to $N[e^{2\phi},\lambda_\phi]$ is at $O(m^2)$, i.e. the linear order of
$\lambda$.

\subsection{Conventions}

Let us first summarize the conventions in this appendix.
Consider a free scalar in a two dimensional space with the metric
\be
ds^2=e^{2\phi(x)}(dx_1^2+dx_2^2),  \label{met}
\ee
which is also written as $g_{ab}=e^{2\phi}\delta_{ab}$. We set the reference metric $g_0$
to be the flat metric $\phi=0$.

The standard free massive scalar action is given by
\be
S_{scalar}=\int dx^2 \s{g}\left[g^{ab}\de_a\vp\de_b\vp+m^2\vp^2\right].  \label{mass}
\ee
We define the Laplacian as
\be
\Delta_g=-e^{-2\phi}\de_a\de_a.
\ee
Note also that the flat space Laplacian is expressed as $\Delta_0=-\de_a\de_a$.
The Green function $G(x,y;g)$ is defined by
\be
\Delta_g G(x,y;g)=\delta(x-y)/\s{g}.
\ee
It is useful to consider the limit $x\to y$ of this Green function:
\be
G_R(x;g)=\lim_{y\to x}\left[G(x,x;g)+\frac{1}{2\pi}\log\frac{D_g(x,y)}{L}\right],
\ee
where $L$ is a renormalization scale and $D_g(x,y)$ is the geodesic distance between $x$ and $y$.

Now we introduce the potential $\Phi(x)$ for the metric (\ref{met}):
\be
e^{2\phi}=\frac{A}{A_0}-\frac{1}{2}A\Delta_0\Phi,   \label{for1}
\ee
where $A$ and $A_0$ are the areas for the metric $g$ and $g_0=1$, respectively.

It is useful to introduce Aubin-Yau action
\be
S_{AY}(g_0,\Phi)=-\int dx^2 \left[\frac{1}{4}\Phi\Delta_0\Phi-\frac{\Phi}{A_0}\right]. \label{for2}
\ee

As shown as eq.(3.40) in \cite{FKZ}, we find the helpful relation:
\be
G_R(x;g)-G_R(x;g_0)=\Phi(x)+\frac{\phi(x)}{2\pi}-S_{AY}(g_0,\Phi). \label{for3}
\ee

\subsection{Constant Mass}

We assume that the mass $m$ in (\ref{mass}) is a non-zero constant.
We can evaluate the partition function $Z_m$ as follows
\ba
-\log Z_m&=&\frac{1}{2}\log \det(\Delta+m^2) \no
&\simeq & \frac{1}{2}\mbox{Tr}\log \Delta + \frac{m^2}{2}\mbox{Tr}[ \Delta^{-1}]+O(m^4).
\ea

Therefore the first order massive correction is given by the difference:
\be
\Delta_{m^2} S=\frac{m^2}{2}\int dx^2 \s{g} G_{R}(x;g)-\frac{m^2}{2}\int dx^2 G_{R}(x;g_0).
\ee

As shown in \cite{FKZ}, we find (by using (\ref{for1}),(\ref{for2}),(\ref{for3}))
\ba
&& \frac{1}{A}\int dx^2\s{g}G_R(x;g)-\frac{1}{A_0}\int dx^2 G_R(x;g_0) \no
&&=-\frac{1}{4}\int dx^2 \Phi\Delta_0\Phi +\frac{1}{2\pi A}\int dx^2\phi e^{2\phi}
-\frac{1}{2}\int dx^2\Phi \Delta_0 G_R(x;g_0)\no
&&\equiv \frac{1}{8\pi}S_M(g;g_0),
\ea
where the final action is called Mabuchi action.

In this way we can evaluate $\Delta_{m^2} S$ as follows \cite{FKZ}
\ba
\Delta_{m^2} S=\frac{m^2A}{16\pi}S_M(g;g_0)+\frac{m^2}{2}(A-A_0)\int dx^2G_{R}(x;g_0)
+\frac{m^2}{4\pi}(A-A_0)\log L,
\ea
where the final term comes from the renormalization.

\subsection{Path-integral Optimization: Position Dependent Mass}

For our purpose of path-integral optimizations,  the mass is chosen to depend on the space coordinate such that (as in (\ref{massfs}))
\be
m^2=m_0^2 e^{-2\phi(x)},
\ee
so that the action is invariant under the scale transformation.
In this case we find that the first order change of effective action is given by
\be
\Delta_{m_0^2} S=\frac{m_0^2}{2}\int dx^2 G_{R}(x;g)-\frac{m_0^2}{2}\int dx^2 G_{R}(x;g_0).
\ee

By using (\ref{for1}),(\ref{for2}),(\ref{for3}), we find
\be
\Delta_{m_0^2} S=\frac{m_0^2}{2}\int dx^2\left[\frac{\phi(x)}{2\pi}-\frac{1}{4}\Phi\Delta_0\Phi\right].
\ee
Note that $\log L$ term cancels out in the above difference.

In this way, for the position dependent mass perturbation, the path-integral optimization can be denoted by minimizing the action (notice the relative sign)
\be
N[e^{2\phi},\lambda_\phi]=S_L-\Delta_{m_0^2} S=
\int dx^2 \left[\de_a\phi\de_a\phi+e^{2\phi}\right]-\frac{m_0^2}{2}\int dx^2\left[\frac{\phi(x)}{2\pi}-\frac{1}{4}\Phi\Delta_0\Phi\right]. \label{totS}
\ee

The full equation of motion is given by the following non local expression

\be
\partial_a\partial_a\phi(x)=e^{2\phi(x)}\Big( 1-\f{m_0^2}{2A}(\Phi(x)-\f{1}{A}\int d^2y\Phi(y)e^{2\phi(y)})\Big)-\f{m_0^2}{8\pi}.
\ee

Let us solve this equation of motion perturbatively in $m_0^2$.
We expand
\be
e^{2\phi}=\f{1}{z^2}+m_0^2\f{2f(z)}{z}
\ee
assuming background metric is Poincare patch.
Expanding $\Phi=\Phi^{(0)}+m_0^2\Phi^{(1)}+\cdots$ and $A=A^{(0)}+m_0^2A^{(1)}+\cdots$, we have
\be
\partial_a\partial_a\Phi^{(0)}=\f{1}{A_0}-\f{1}{zA^{(0)}}
\ee
so we have
\be
\Phi^{(0)}=\f{1}{2A_0}z^2+Cz-\f{z\log z}{A^{(0)}}+D.
\ee
Assuming boundary of AdS has infinite volume, and C and D are order $\mathcal{O}(vol^0)$,
then we can simplify the equation of motion significantly

\be
(zf)''=\f{2f}{z}-\f{1}{8\pi}.
\ee

The general solution is
\be
f(z)=-\f{1}{24\pi}z~\log\f{z}{a},
\ee
where $a$ is an arbitrary positive constant. By imposing boundary condition
\be
e^{2\phi}|_{z=\ep}=\f{1}{\ep^2},
\ee
we can fix $a=\ep$.
Therefore the perturbative solution is
\be
e^{2\phi}=\f{1}{z^2}\left(1-\f{m_0^2}{12\pi}z^2\log\f{z}{\ep}+\mathcal{O}(m_0^4)\right).
\ee
It clearly shows that the space is squeezed compared with the massless case $m_0=0$.
Let us estimate our path-integral complexity. For simplicity we assume integration range of our path-integral
 complexity can be limited to the range
 \be
 \ep\leq z\leq \ti{a}
 \ee
where $\ti{a}$ is defined by the condition
\be
e^{2\phi}|_{z=\ti{a}}=0.
\ee
In the first order approximation in $m_0^2$, we have
\be
1=\f{m_0^2}{12\pi}{\ti a}^2\log\f{\ti a}{\ep}.
\ee
Using these assumptions, we can see our path-integral complexity indeed decreases.
The change of our path-integral complexity is
\be
\Delta C=-\f{m_0^2 \ti{a} L}{12\pi}(3+{\rm log}\f{1}{\ep\ti{a}^2}).
\ee
This is negative, when $\f{1}{\ep\ti{a}^2}$ is large enough. This is realized for example when $m_0^2=\mathcal{O}(\ep^2)$.
Although these results are expected behavior of metric and path-integral complexity, it is difficult to compare the above behavior with the AdS/CFT because we are considering
a free scalar field theory, which is not holographic and because the mass perturbation is not
a primary operator in two dimensions.

\end{document}